\documentclass[amsmath,amssymb,showpacs,twocolumn]{revtex4}
\usepackage{amssymb}
\def\ket{\rangle}
\def\bra{\langle}

\usepackage{graphicx}
\usepackage{dcolumn}
\usepackage{bm}

\begin{document}
\title{Multiple Entropy Measures for Multipartite Quantum Entanglement }
\author{Dan Liu }
\author{ Xin Zhao}
\author{Gui Lu Long  \footnote{E-mail:
gllong@mail.tsinghua.edu.cn}} \affiliation{
Key Laboratory for Atomic and Molecular NanoSciences and Department of Physics, Tsinghua University, Beijing 100084, China; \\
Tsinghua National Laboratory for Information Science and Technology, Tsinghua
University, Beijing 100084
}%
\date{May 26, 2007}

\begin{abstract}
A new entanglement measure, the multiple entropy measures (MEMS), is proposed to
quantify quantum entanglement of multipartite quantum state. The MEMS is vector-like
with $m=[N/2]$, the integer part of $N/2$, components: $[S_1,\; S_2,\cdots, S_m]$, and
the $i$-th component $S_i$ is the geometric mean of $i$-body partial entropy of the
system. The $S_i$  measures how strong an arbitrary $i$ bodies from the system are
entangled with the rest of the system. The MEMS is not only transparent in physical
picture, but also simple to calculate. It satisfies the conditions for a good
entanglement measure. We have analyzed the entanglement properties of the GHZ-state,
the W-states and cluster-states under MEMS. The cluster-state is more entangled than
the GHZ-state and W-state under MEMS.
\end{abstract}
 \maketitle

Entanglement is an exciting property of quantum mechanics, and it reflects the property
that a quantum system can simultaneously appear in two or more different states
\cite{schoedinger,einstein}. It plays an essential role in quantum information and
computation \cite{Nielsen}. The quantification of entanglement is a challenging topic,
many works have been contributed to it
\cite{r4,r5,r6,r7,r8,r9,r10,r11,r12,r13,r14,r15,r16,r17,r18,r19,r20,r21,r22,r23,r24,r25,r26,r27,r27p,r27p2}.
Entanglement measures such as  the partial von Neumann entropy \cite{r4,r5}, the
elative entropy \cite{r6,r7}, entanglement of formation \cite{r4,r5},  concurrence
\cite{r8,r12,r19,r24} have been proposed and studied extensively over the years.

Basic requirements for defining entanglement measure  are \cite{r7,r11}: (1) it is a
positive map of the density matrix; (2) for separable state it is zero; (3) it is
invariant under local unitary operation and does not increase under local operation and
classical communication (LOCC); (4) for pure bi-partite system, it reduces to the
partial entropy measure. Any measure that satisfies conditions (1-3) is said to be an
entanglement monotone. In practice, some proposed measures do not satisfy all these
conditions, though they are still useful for practical application. Most of the
entanglement measures suffer from one or another difficulty in extending to multi-qubit
system or are infeasible to manipulate. It has been realized that it is not enough to
use just a single quantity to measure entanglement. Recently, a universal entanglement
measure was proposed by Partovi \cite{r22}, and it uses $N!/2$ quantities for an
$N$-qubit system. When $N$ increases the number of quantities increases rapidly and it
becomes intractable.

Entanglement measure for two-qubit pure states has been solved completely. It can be
measured by the partial entropy of entanglement, and all other measures  are equivalent
to it. Can we use partial entropy to measure multipartite quantum entanglement?  When
$N=3$, a complete set of reduced density matrices is
$\{(\rho_\psi)_{AB},(\rho_\psi)_{BC},(\rho_\psi)_{AC},(\rho_\psi)_{A},
(\rho_\psi)_{B},(\rho_\psi)_{C}\}$, where $A$, $B$, $C$ refer to the three qubits
respectively. Pan et al used the arithmetic average entropy of the individual single
reduced density matrices as the measure of entanglement, and studied the extremal
entangled state for three-qubit pure state\cite{r23}. It has been found that there are
three types of extremal entangled states \cite{r23}, namely the GHZ state, the W state
 and another entangled state with average entropy $0.707$.
  Cao et al. \cite{r25} proposed to use the entropy product as an entanglement
measure, and it is smoothly varying. Thus it has been shown that for three-qubit
quantum system, a modified partial entropy, the `entropy product' given in Ref.
\cite{r25}
\begin{eqnarray}
S_E=\prod_{i=1}^N E_i\label{es1},
\end{eqnarray}
where\begin{eqnarray} E_i=-{\rm Tr}[(\rho_\psi)_i\log_2 (\rho_\psi)_i]\label{eei}
\end{eqnarray} gives a good measure of the entanglement of the three-qubit
quantum system.

\begin{figure}
\includegraphics[width=8cm]{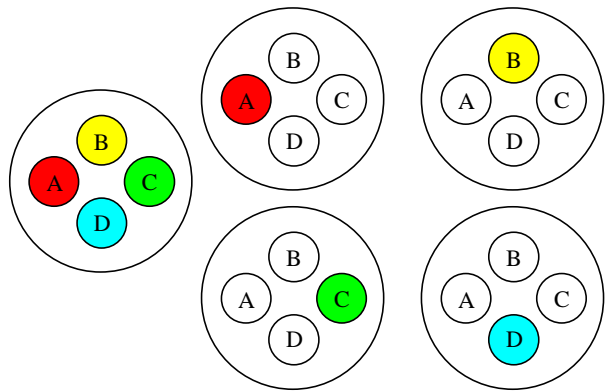}
\caption{(color online)Schematic illustration. On left is a pure state with qubit A, B,
C, D. $S_1$ is the geometric mean of the four cases on the right, where each large
circle represents the partial entropy of a qubit with the rest three qubits being
traced out.}\label{f1}
\end{figure}

However for pure states with $N\geq 4$ qubits, it is not enough to describe the
entanglement properties using only $S_E$. $S_E$ characterizes the  relation of one
single qubit with the other three qubits only, as shown in Fig.\ref{f1}. The
entanglement between any chosen two qubits with the other two qubits is not considered
at all. For instance, a four-qubit state with two EPR pairs gives $S_E=1$, the same as
that of a GHZ-state. However, it is not a genuine entangled state. In fact, it is
necessary to consider the entropy of two qubits for four-qubit pure states. Of course,
 Higuchi and
Sudbery proposed to use the average mean of two-particle entropies as a measure of
entanglement, and studied four-qubit entangled state with respect to it \cite{hspla}.

For multipartite pure states, multi-partite entropies of reduced many-body density
matrices should be considered in order to describe the entanglement. According to the
result in Ref. \cite{r25}, it is better to use geometric mean rather average mean
entropies so that discontinuity can be avoided.  Based on this analysis, we propose an
entanglement measure, the multiple entropy measures (MEMS) for multipartite pure
quantum states,
\begin{eqnarray}
S_1&=&\left[\prod_i E_{i}\right]^{1\over N},\\
 S_2&=&\left[\prod_{i,j=1}^N E_{i,j}\right]^{1\over C_N^2},
 \label{s2}\\
S_3&=&\left[\prod_{i,j,k=1}^N E_{i,j,k}\right]^{1\over C_N^3},
 \label{s3}\\
&&\cdots \nonumber\\
S_n&=&\left[\prod_{i_1 i_2 ...i_n}^N E_{i_1,i_2,\cdots,i_n}\right]^{1\over C_N^n},
\end{eqnarray}
where
\begin{eqnarray} E_{i_1i_2\cdots i_n}=-{\rm Tr}[(\rho_{\Psi})_{i_1\cdots
i_n}\log_2 (\rho_{\Psi})_{i_1\cdots i_n}]
 \label{xiE}
\end{eqnarray}
is the reduced von Neumann entropy for the $i_1$-th, $\cdots$ $i_n$-th qubit with the
other $N-n$ particles being traced out ($i_1,\cdots i_n=1,2,\cdots ,N$ and the $i$'s
are not equal to each other), and \begin{eqnarray} C_N^n={N!\over (N-n)! n!},
\end{eqnarray} is the number of possible $n$ qubits reduced density matrices. The $S_i$
is thus the geometric average entropy of $i$ qubits reduced density matrices.
 It should be emphasized that $n\le N/2$, namely we need only consider $[N/2]$ number of
 reduced entropies,
 because  $S_n=S_{N-n}$ due to the
Schmidt decomposition theorem.

The physical picture for $S_i$ is simple and clear. $S_1$ manifests the entanglement
feature in terms of single particles, as shown in Fig.\ref{f1}. If one qubit is
disentangled from the rest, then its reduced density matrix has zero entropy and hence
makes $S_1$ zero. Among the $N$ qubits, if one qubit, say $A$,  is in an entangled
environment, then $S(\rho_A)=1$. If all the qubits are in such an environment, then
$S_1$ has a maximum value of 1. Similarly, for $S_i$, if every $i$ qubits in the system
feel a maximally entangled environment, then $S_i$ will have a value of $i$. For
instance, the maximum value of $S_2$ is 2, and $S_3$ is 3 and so on.

\begin{figure}
\includegraphics[width=8cm]{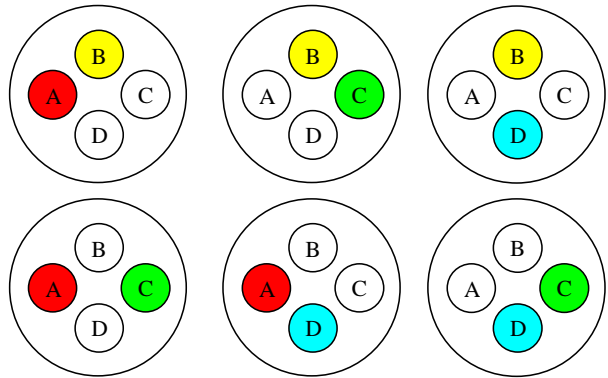}
\caption{(color online)Schematic illustration of $S_2$ of four-qubit system. $S_2$ is
the geometric mean of the six cases.}\label{f2}
\end{figure}

Now we show that the MEMS satisfies the requirements for a good entanglement measure.
First it is apparent that $S_i$ is a positive map of the density matrix. Secondly, we
notice that for an $N$-qubit product state, all the $S_i$'s are zero, where $i$ ranges
from 1 to $m=[N/2]$. Thirdly, each of the entropy in the product is invariant under
local unitary operation, and their products are also invariant under local unitary
operation. The entropies do not increase under LOCC, and hence all the $S_i$'s do not
increase under LOCC. Fourthly, for pure bi-partite pure state, it reduces to the
partial entropy.  Thus the MEMS is a good measure for entanglement.

What interests us most is the genuine entangled states where all the $S_i$'s are
nonzero. If one of the $S_i$ is zero,  at least one group of $i$ qubits, viewed as a
single object, is disentangled from the remaining $N-i$ qubits. Thus it is not a
genuine $N$ qubits entangled state. For example, the product of two EPR pairs is not a
genuine entangled state because $S_2=0$ though $S_1=1$.

Ideally a maximally entangled $N$-qubit state is one that maximizes all the $S_i$'s,
namely $S_1=1$, $S_2=2$, $S_3=3$,$\cdots$, $S_{[N/2]}=[N/2]$. For $N=2$, the Bell-state
is a maximally entangled state because it has $S_1=1$. For $N=3$, the GHZ-state is a
maximally entangled state because $S_1=1$.

For $N=4$,  Higuchi and Sudbery have shown that such state does not exist, namely $S_2$
can not be saturated by 4-qubit system. They have shown that four four-level particle
system, there exists state whose two-particle reduced density matrices are completely
mixed \cite{hspla}. Thus $S_2$ can be saturated for such state. It is not yet known if
more qubit system state can saturate $S_2$. According to Ref. \cite{hspla,hs2}, the
local maximum entangled state with respect to the average mean of two-qubit entropies,
and also with respect to $S_2$, is the $M_4$ state. The $S_2$ for the $M_4$ state is
$(1+{1\over 2}\log_2 3)\approx 1.7925$.  We shall see next that the GHZ-state, the
W-state and the cluster state are not such $S_i$ saturated entangled states.

It is easy to show that for an $N$-qubit GHZ state all the $S_i$'s has a value of 1. It
is the maximum for $S_1$, but is not for other $S_i$'s. It is maximally entangled with
respect to $S_1$, namely each individual qubit feels maximally entangled. However, for
a group of $i$ qubits inside the system, it  still feels as if it were a qubit, and has
$S_i$=1.

\begin{figure}
\includegraphics[width=8cm]{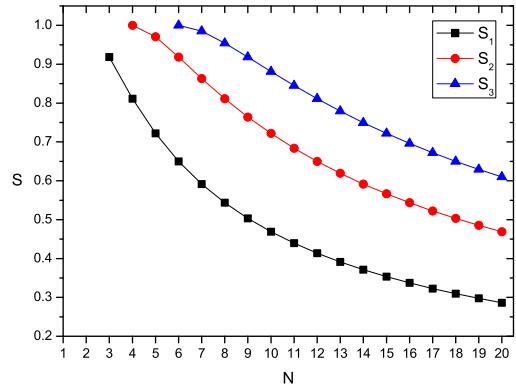}
\caption{(color online)$S_1$, $S_2$, $S_3$ versus $N$ for W-state.}\label{f3}
\end{figure}

\begin{figure}
\includegraphics[width=8cm]{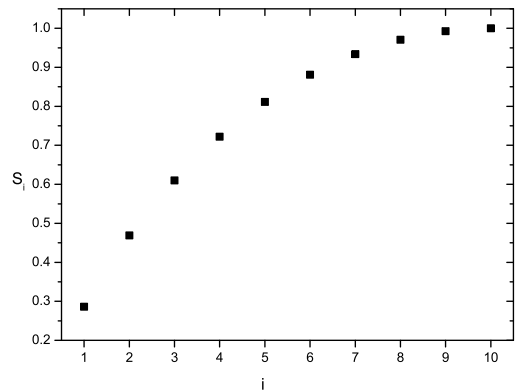}
\caption{ $S_i$ versus $i$ for $N=20$ for the W-state.}\label{f4}
\end{figure}

The W-state,
\begin{eqnarray}
|W_N\ket=\sqrt{1\over N}\left\{|0\cdots 01\ket +\cdots+|10\cdots0\ket\right\},
\end{eqnarray} is another class of entangled state. After some
tedious calculation, we have  for the W-state
\begin{eqnarray}
S_1&=&\left[-{N-1\over N} \log_2 {N-1 \over N}-{1\over N}\log_2{1\over N}\right],
\\
S_2&=&\left[-{N-2\over N} \log_2 {N-2 \over N}-{2\over N}\log_2{2\over
N}\right]\\
S_3&=&\left[-{N-3\over N} \log_2 {N-3 \over N}-{3\over N}\log_2{3\over
N}\right]\\
\cdots\nonumber\\
S_i&=&\left[-{N-i\over N} \log_2 {N-i \over N}-{i\over N}\log_2{i\over
N}\right]\\
\cdots,\nonumber
\end{eqnarray}
It is interesting to see that all the $S_i$'s are less than 1. In Fig.\ref{f3}, we plot
$S_1$, $S_2$ and $S_3$ against $N$, and we see except one point in which $(N-i)/2=1/2$,
all the $S_i$ are less than 1.  In Fig.\ref{f4}, we have plotted $S_i$ for $i=1,2,
[N/2]$ for a W-state with $N=20$. It is usually less than 1. Hence in general the
GHZ-state is more entangled than the W-state in terms MEMS.

Now it is interesting to examine the cluster state, which has been experimentally
realized recently for the four-qubit system \cite{r29}. It is found that for this
state, $S_1$ is always 1. Hence it is maximally entangled with respect to $S_1$.  In
this context, the cluster state is as entangled as the GHZ-state. But for a general
$N$-qubit cluster state
\begin{eqnarray}
S_2=2^{(C_N^2-2)\over C_N^2},\label{ecluster2}
\end{eqnarray}
which is slightly less than 2. The reason can be seen clearly from this example,
\begin{eqnarray}
|\Phi_4\rangle={1\over 2}(|0000\rangle+|0011\rangle+|1100\rangle-|1111\rangle).
\end{eqnarray}
Naming the qubits as A, B, C and D in order. There are altogether 6 different ways to
choose two qubits. For the case of AB, or CD, 00 and 11 always come together just like
one qubit, hence the reduced density matrices contribute 1 to the product. Other four
cases contribute 2 to the product each. Thus the $S_2$ for a 4-qubit cluster state is
$2^{2/3}=1.587$.

For cluster state with $N=6$, $ S_3=\sqrt[20]{2^{10}3^8}\simeq 2.195$, which mean that
two clusters see an entanglement of a  two-level system, and contribute to 1's to
$S_3$, and 10 such three-qubit clusters each contributes 2, and 8 such three-qubit
clusters each contributes 3 to the product. This state is apparent more entangled than
a 6-qubit GHZ state.

It is interesting to examine the transformation from one state to another state. From
the properties of MEMS, we have the following no-go theorem:

{\noindent\bf Theorem}:  $N$-qubit states $|\psi\ket$ cannot be transformed into
$|\psi'\ket$  if they do not satisfy $S_i\ge S_i'$ for all $i$'s.

{\noindent\bf Proof}: Since MEMS does not increase under LOCC,  any LOCC on state
$|\psi\ket$ cannot increase the value of $S_i$ so as to make $S_i\ge S_i'$.

Notice that this no-go theorem provides a necessary, not a sufficient condition for
transformation of states using LOCC. One example is the GHZ-state and W-state. It is
true that it is impossible to transform W-state to GHZ-state using LOCC. But it is also
impossible to transform from GHZ-state to W-state. It is possible to transform from a
cluster state to either GHZ-state or W-state, which is not forbidden by our theorem.

The MEMS can be extended directly to quantum systems with particles of different
dimensions. The corresponding maximum value of the $S_i$ should vary accordingly. For
instance $S_1=\sqrt{2 \times 4}$ for a two-body system of 2 level and 4 level quantum
systems.

Now we turn to mixed states. As pointed out in Ref. \cite{r30} that there are different
definitions of mixed states. A proper mixed state is the average state of a
hypothetical particle from an ensemble with $N_i$ particles in state $|\psi_i\ket$, and
its density matrix is
\begin{eqnarray}
\rho=\sum_i p_i|\psi_i\ket\bra \psi_i|,
\end{eqnarray}
where $p_i=N_i/N$, and $N$ is the total number of particles in the ensemble. The
averaged MEMS for the proper mixed state is
\begin{eqnarray}
S_i=\sum_j p_j S_i(j),
\end{eqnarray}
where $S_i(j)$ is the $S_i$ for pure state $|\psi_j\ket$. For improper states, the MEMS
will be defined as the simultaneous minimum value of $S_i$  over all possible
decompositions of $\rho$.

In summary, a new entanglement measure, the MEMS,  has been proposed and investigated.
This measure satisfies the criteria of a good entanglement measure. It has a simple and
clear physical interpretation, and is easy to calculate. Some well-known entangled
states have been studied under MEMS. It will be interesting to study further the
properties of MEMS.

We thank Prof. S. M. Fei for helpful discussion. Work is supported by the National
Natural Science Foundation of China, Grant No. 10325521, 60433050(GLL)  60635040, the
973 program 2006CB921106 (XZ),  the SRFDP program of Education Ministry of China, No.
20060003048 (DL).

\end{document}